# Recycling an anechoic pre-trained speech separation deep neural network for binaural dereverberation of a single source

Sania Gul[1], Muhammad Salman Khan[2*], Syed Waqar Shah[1], Ata Ur-Rehman[3]

[1]Department of Electrical Engineering, University of Engineering and Technology, Peshawar, Pakistan.

[2]Department of Electrical Engineering, College of Engineering, Qatar University, Doha, Qatar.

[3]Department of Electrical Engineering, MCS, NUST, Islamabad, Pakistan.

**Abstract_** *Reverberation results in reduced intelligibility for both normal and hearing-impaired listeners. This paper presents a novel psychoacoustic approach of dereverberation of a single speech source by recycling a pre-trained binaural anechoic speech separation neural network. As training the deep neural network (DNN) is a lengthy and computationally expensive process, the advantage of using a pre-trained separation network for dereverberation is that the network does not need to be retrained, saving both time and computational resources. The interaural cues of a reverberant source are given to this pre-trained neural network to discriminate between the direct path signal and the reverberant speech. The results show an average improvement of 1.3% in signal intelligibility, 0.83 dB in SRMR (signal to reverberation energy ratio) and 0.16 points in perceptual evaluation of speech quality (PESQ) over other state-of-the-art signal processing dereverberation algorithms and 14% in intelligibility and 0.35 points in quality over orthogonal matching pursuit with spectral subtraction (OSS), a machine learning based dereverberation algorithm.*

1. # Introduction

When sound is emitted from a source inside an enclosure, it reaches the human ear both by the direct path and by the reflections from the walls and the surrounding objects. These reflections are called 'echoes' or 'reverberations'. Reverberation is a natural phenomenon unavoidable in an enclosed environment. In noiseless enclosures, reverberations degrade the speech intelligibility for hearing impaired and non-native listeners. In noisy enclosures, reverberations smear the speech intelligibility even for the normal hearing listeners [1].

For machine listening, reverberation results in degradation of the output performance of automatic speech recognition (ASR) systems, automatic speaker verification (ASV), speech separation systems, hands-free mobile communication, teleconferencing and so on. Reverberations are more problematic than noise, as the noise is additive in nature while reverberations are convolutive [2].

The emissions from a source are convolved with the room impulse response (RIRs), which are characterized by the path; an individual emission takes to reach the listener. The sound waves reaching through the direct path are the first ones to reach the listener. The sound waves which reach the listener within 50ms after the direct sound are regarded as the early reverberations and those reaching after 50ms are classified as late reverberations [3]. The RIR break down in direct, early and late reflections is shown in figure 1(a) below.

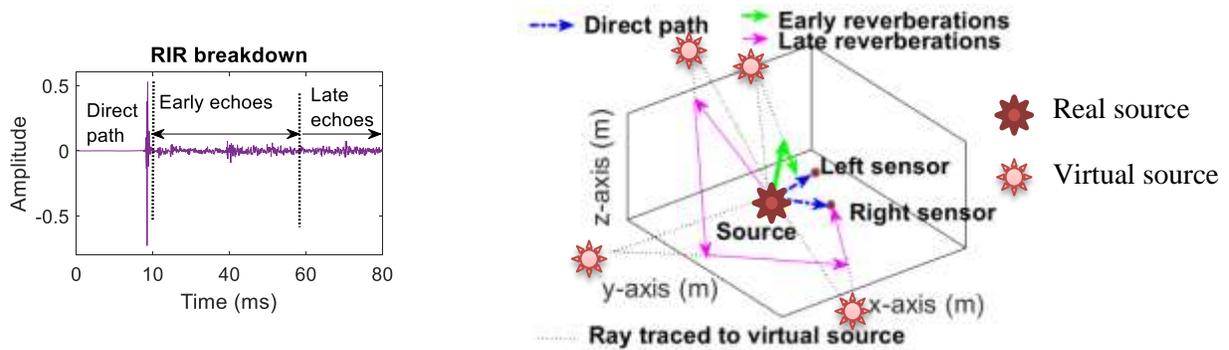

**Figure 1:** (a) RIR breakdown in early and late reflections. (b) Creation of virtual sources by reverberations

While, early reverberations help in increasing the speech intelligibility by effectively amplifying the sound amplitude, the late reflections are detrimental to speech perception, source separation and sound localization [4]. Both early and late reflections reduce the accuracy of source localization, perceived source width and the ability to detect and understand one source in the presence of other competing sources [5]. The reverberation causes temporal smearing of speech energy and thus detrimental to automatic speech recognition (ASR) performance [6]. The smearing effect rises with the increase in reverberation time ($RT_{60}$) as depicted in Figure 2.

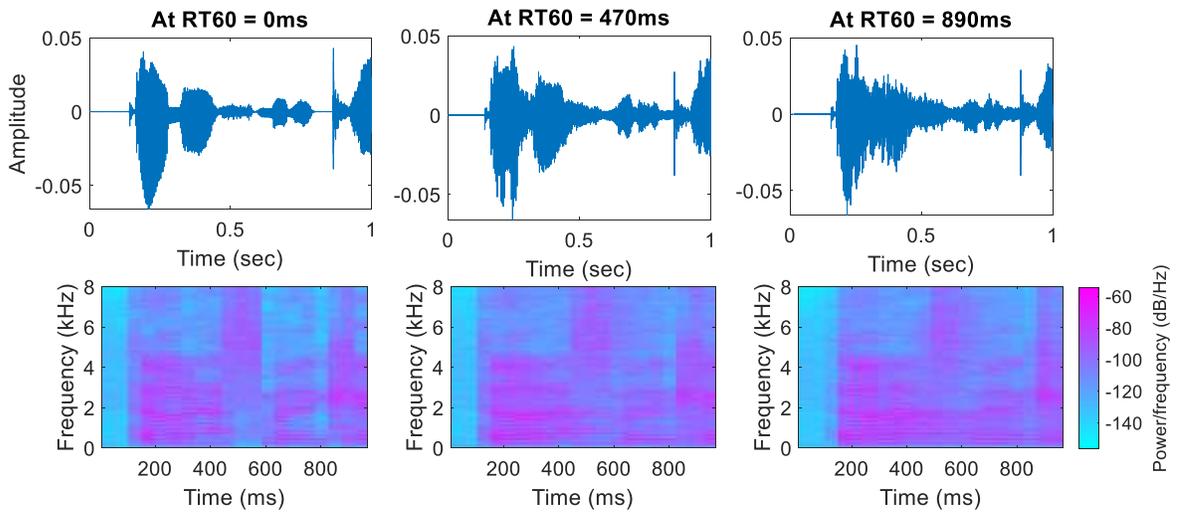

**Figure 2:** Increase in energy smearing with the increase in *RT$_{60}$*.

Reverberations are the delayed and decayed versions of the original source signal [2]. Apart from their reduced strength and delayed arrival, reverberations can also be characterized by their interaural cues. As the reverberations can be viewed as the same source signal coming from several different sources placed at different locations in an enclosure [7], the spatial cues of reverberations are different from those of direct waves [8]. These reflections act as virtual sources, with interaural cues different from the original source generating them [9] as shown in figure 1(b).

In past, many techniques have been developed to combat the fatal effects of reverberations, including signal processing techniques (e.g. [10], [11], [12] and [13]) (commonly called as 'classical' methods), psychoacoustic methods (e.g. [4] and [14]), machine learning algorithms (e.g. [15]), and now the latest trend is the use of DNNs (e.g. [2] using the BLSTM (bidirectional long short term memory ) and [16] using the GAN ( generative adversarial networks)).

In [12], the spectral components of late reverberations are subtracted from the reverberated signal to obtain the dereverberated speech; hence the method is named as 'spectral subtraction (SS)'. Another popular classical method to dereverberate a signal is 'inverse filtering' where the impulse response of filter (i.e. the room impulse response) is estimated from the reverberated speech itself and the inverse of this estimated response is then applied on the reverberated speech to obtain the echo-free speech. In spectral magnitude inverse filtering (SMIF) model of [1], instead of estimating the filter response in time domain, the frequency domain response, called convolutive transfer function (CTF) is estimated. Instead of inversing this complex valued CTF, only the inverse filters of CTF magnitude are formulated and applied on the STFT (short time Fourier transform) magnitude of reverberated signal to dereverberate it. Another effective signal processing dereverberation technique is 'filter shortening'. The reverberation causes the lengthening of RIR as shown in figure 1 (a) (the portion of the response belonging to early and late echoes). The weighted prediction error (WPE) algorithm proposed in [13], shortens this impulse response by using the sub-band domain multi-channel linear prediction filters. Using these filters, the RIR shortening process is generalized and can work in all kinds of acoustic conditions.

In [14], the psychoacoustic 'precedence effect (PE)' is implemented for dereverberation. The precedence effect is an auditory mechanism that aids humans in localizing sounds in reverberant environments by giving more perceptual weight to the direct sound as compared to the late reflections. The PE is achieved in [14] by decomposing the speech band into psychoacoustically significant channels using gammatone filter bank and preserving only the initial wave fronts of each channel, as it is observed that the initial wave fronts are not contaminated by late reflections. The clean speech is then reconstituted by adding back all the channels.

In orthogonal matching pursuit with spectral subtraction (OSS) model proposed in [15], machine learning K-SVD (K- singular value decomposition) algorithm, a generalization of K-means algorithm is used for dereverberation. It is a two stage process [15]. In the first stage, orthogonal matching pursuit on the orthogonal bases learned by K-SVD is used to obtain a sparser representation of reverberated signal. The reverberation time of the room is estimated by the recorded hand clap and is used to create a time-domain envelope. In the second stage, this time domain envelope is used to estimate the energy of the late reverberations which is then removed by spectral subtraction.

The model proposed in [2] is a DNN based dereverberation model accompanied by beamforming and trained on the real and imaginary (RI) spectral components of the direct path signal. Due to the extremely high learning capabilities of neural network and the use of beamformer, the model performs extremely well in forensic speech applications e.g. ASR (Automatic speech recognition) and ASV (Automatic speaker verification). In [16], GAN is trained on log-power spectra (LPS) of clean speech. It is observed that training the GAN on LPS features instead of MFCC (Mel-frequency cepstral coefficients) results in reduced CER (character error rate), when used as a front end for ASR systems. It has been shown in [2] and [16] that dereverberation by deep learning has produced better results in terms of speech quality and intelligibility as well as reduced word error rate (WER) in automatic speech recognition applications in comparison to the signal processing and machine learning techniques. However, the biggest downside of deep learning models is their high computational cost, lengthy training durations and the requirement of large training and testing datasets. Although our proposed dereverberation algorithm is a deep learning model, it is different from the deep learning models described above as it does not require any training, yet it produces better results than the state-of-the-art signal processing and machine learning models discussed above.

In this paper, we will use the differences in the interaural cues of the direct path and reverberations to dereverberate a speech source. For this purpose, we have selected a pre-trained speech separation DNN, which was trained on the interaural cues of a single source placed in anechoic conditions. To the best of our knowledge, this is the first time that any pre-trained speech separation neural network, is being recycled for the task of dereverberation of a single speech source without any need of further training or modification in the pre-trained speech separation network. The speech enhancement GAN (SEGAN) [17] (originally designed for the denoising application) was tested for speech dereverberation in [16], as the front-end in the automatic speech recognition (ASR) application. But the model (without any structural modifications) failed to meet the desired objective of reducing the character error rate (CER), even after retraining on the dataset of the new domain. It is believed that time domain enhancement used by SEGAN is only effective for the noise removal, but not for dereverberation.

The rest of the paper is organized as following. In the section 2, we focus on related work, while we provide the overview of our proposed system in section 3. In section 4, we will describe the experimental setup, the evaluation criteria and the comparison algorithms. We will present experimental results and comparison statistics of different models in section 5 and conclude the paper in section 6.

## 2. Related Work

We will briefly describe here the pre-trained speech separation neural network model 'SONET' [18] that we have incorporated in our proposed dereverberation algorithm.

'SONET' [18] is a binaural anechoic speech separation model, trained on time frequency (TF) interaural spectrograms, of two spatially separated sources. Although, SONET is source separation model, it was never trained on speech mixtures. Rather, its training dataset comprised of two classes. Each class consists of spatial spectrograms produced by the direct path of a single active source. It is only after the training, that the SONET was exposed to a speech mixture (i.e. both sources active simultaneously). The core network of [18] is U-Net, which is deep learning convolutional neural network architecture. Two U-Nets were trained, the first one on the interaural level difference (ILD) spectrograms (ILD-SONET), and the second one on the interaural phase difference (IPD) spectrograms (IPD-SONET). During training phase, each U-Net was trained from the dataset containing the interaural spectrograms belonging to two classes, the target and the masker. Every training sample of each class was accompanied by its corresponding ground truth (GT) mask. GT of target was an image consisting of all white pixels (pixel value =255), while that of masker was an image of all black pixels (pixel value = 0). The size of GT image was equal to the size of the ILD/IPD spectrogram. After training, the spectrograms of audio mixture (containing both target and masker active simultaneously) were presented to U-Nets for clustering the audio mixture's TF units into individual classes.

The motivation behind recycling a pre-trained speech separation model SONET [18] for the dereverberation task is that, SONET had already learnt the direct path interaural cues of a single source during its training phase, when this source is the only active source inside an anechoic room, and would not require any additional training or change in its structure, if it were to be used for Dereverb-SONET (D-S) model. It is anticipated that SONET may still able to recognize efficiently the direct path cues in the presence of newly created cues generated by reverberations, when the same active source for which SONET has already learnt the direct path cues, is now placed in an echoic room.

## 3. Proposed Methodology

We call our dereverberation algorithm as 'Dereverb-SONET (D-S)': as it is a single speech binaural dereverberation model based on SONET. The D-S employs recycling concept, where a DNN trained for one purpose (in our case separation) is used for another purpose (i.e. dereverberation of a single speech source), without the need to retrain the existing network or change any of its layer. This concept is different from transfer learning, which reuses a pre-trained network for a new application, but requires slight modifications in the existing network's output layers and retraining with a small dataset for fine tuning the pre-trained network parameters for the new application.

When a speech source is placed inside a reverberant enclosure, the signals emitted from it would reach the listener both by the direct path and by the reflections off the walls and the surroundings objects. SONET was trained on the interaural spectrograms of the direct path signals of two spatially separated sources in an anechoic room. It separated the two sources by using the differences in their interaural cues. So, when this trained model is used for the dereverberation task of a single source, it uses the same scheme of discrimination of interaural cues, as the direct path cues depicts the location of real source while the reverberation cues depict the position of virtual sources, shown in figure 1 (b). These two types of waves (the direct path emitted by the real source, and the reflected waves emitted by the virtual sources) have differentiable interaural cues and are separable by SONET.

In our proposed methodology, speech from the active source $s_i$ is collected at the two microphones placed as shown in figure 3. Each microphone signal $x_k$, is given as in eq. (1)

$$x_k(p) = s_i(p) * h_{ki}(p) \qquad for\ k = \{1,2\} \qquad (1)$$

Where $x_k$ is the signal collected at the $k^{th}$ microphone, $h_{ki}$ represents the RIR between the source $s_i$ and the $k^{th}$ microphone, '*' represents the convolution operation and $p$ represents the discrete time index, when the sampling frequency, at which microphone signal ($x_k$) samples are taken is $f_s$. The signal notations used in eq. (1) are also shown in figure 3.

STFT of the signal $x_k$ collected at the $k^{th}$ microphone is taken after segmenting $x_k$ into overlapping frames of length $L$ and windowing each frame with the window function $w(p)$ as given in eq. (2).

$$X_k(\omega, m) = \mathcal{F}(w(p)x_k(p)) \qquad (2)$$

Where $\omega$ is the discrete circular frequency index, $m$ is the time frame index, $\mathcal{F}$ is the STFT operator and $w(p)$ is the hamming window function given as $w(p) = 0.54 - 0.46 \cos(2\pi p/N), 0 \leq p \leq N, where\ N = L - 1$, and $L$ is the window length (See table 2 for STFT parameters).

The ratio of the STFT of signals at both microphones is taken to generate the interaural spectrogram as given in eq. (3).

$$\frac{X_1(\omega, m)}{X_2(\omega, m)} = \alpha(\omega, m)e^{i\phi(\omega, m)} \qquad (3)$$

Where $\alpha(\omega, m)$ is the ILD and $\phi(\omega, m)$ is the IPD at a TF point having discrete circular frequency index $\omega$ and time frame index $m$. Converting the ILD to decibels (dB) at each TF point by the formula

$$ILD(dB) = 20\log_{10}\alpha(\omega, m) \qquad (4)$$

The ILD and the IPD spectrograms (obtained from equations (3) and (4) over all TF points) are given as an input to the pre-trained speech separation neural network 'SONET', which will then discriminate between the direct path interaural TF points from the reverberant TF points. As 'SONET' was trained on a two class dataset, so, each SONET (ILD or IPD) will generate two soft masks at its softmax layer. The first soft mask from both SONETs (ILD and IPD) belongs to the direct path and the second soft mask of each SONET belongs to reverberation class. These masks are probabilistic masks (with value at each TF point ranging from 0 to 1), indicating the degree of association of each TF point to the two classes (direct path and reverberations). The direct path ILD and IPD mask are then combined according to eq. (5) to prepare the sub-band mask as

$$Sub-band\ mask\ M(\omega, m)$$
$$= \begin{bmatrix} M_{IPD}(\omega, m), +\omega = 1:96\ and -\omega = 929:1024\ for\ 0 < f < 1.5\ kHz\ ; \\ M_{IPD}(\omega, m) \times M_{ILD}(\omega, m), +\omega = 97:256\ and -\omega = 769:928\ for\ 1.5\ k < f < 4\ kHz; \\ M_{ILD}(\omega, m), +\omega = 257:512\ and -\omega = 513:768\ for\ 4\ k < f < 8\ kHz] \end{bmatrix} (5)$$

where $M_{IPD}(\omega, m)$ is the IPD sub-band mask, and $M_{ILD}(\omega, m)$ is the ILD sub-band mask and $M_{IPD}(\omega, m) \times M_{ILD}(\omega, m)$ is the product mask at the analog frequency $f$ corresponding to discrete frequency $\omega$. The values of $\omega$ shown in eq. (5) corresponds to 1024 discrete point STFT. As STFT produces both positive and negative discrete frequencies $\omega$, so there exists both positive and negative discrete frequency bands corresponding to a unique analog frequency band.

The sub-band mask in eq. (5) is devised according to the strength of spatial cues described in [19]. The IPD cues are strong in the frequency band of 0 to 1.5 kHz, while both ILD and IPD cues are weak in the region between 2 to 4 kHz and above 4 kHz the ILD cues are very much stronger than the IPD cues due to 'head shadow' effect [19]. So, we have developed a sub-band mask for the retrieval of direct-path TF points by taking the portions of the direct path ILD and IPD soft masks according to their strength in different frequency bands as shown in equation 5.

This sub-band mask is applied on the spectrograms of reverberant speech source $X_1(\omega, m)$ and $X_2(\omega, m)$, and the outputs are added together to retrieve the target spectrogram which is then converted from TF domain back to time domain by inverse STFT (ISTFT), and evaluated against the direct path sound (clean speech source convolved with anechoic RIRs of [4]), collected at the left microphone.

## 4. Experimental Evaluation Parameters

This section includes dataset, experimental layout, room impulse responses, comparative algorithms, and the metrics to evaluate the performance of our proposed dereverberation model D-S.

### 4.1. Experimental Setup

Experiments were performed with a single active source $s_i$ in a room as shown in figure 3.

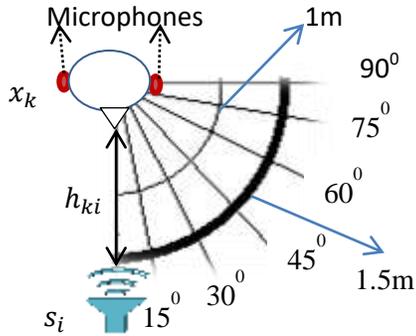

**Figure 3:** Equipment setup. The source is currently placed at $0^o$ from the binaural setup at a distance of 1.5m. The inner circle represents 1m and the outer circle represents 1.5 m spacing between the source and binaural setup.

According to the binaural room impulse responses (BRIRs) of [4] and [5] used in this experiment, the distance between the two microphones is 145 mm, which is the average distance between the two human ears. The distance between the source and the binaural setup is 1.5m when the BRIRs of [4] are used and it is 1m when the BRIRs of [5] are used. The position of source is varied from 0 to 90 degrees in the increment of 15 degrees.

As the pre-trained model 'SONET', is trained for source positions varying from 0 to 90 degrees towards right side), so we have restricted the positions in this range. Any other position, for example, the source placed somewhere near the left microphone or behind the binaural setup, would require retraining the SONET, as the interaural cues generated by such a source would be different from those on which SONET was trained, negating the recycling concept.

However, the proposed dereverberation model is tested both with target-microphone spacing of 1.5 meters (for which the 'SONET' was trained) and with target-microphone spacing of 1 meter (for which SONET was not trained).

The sources are recordings of human speakers and are directional. The BRIRs of [4] and [5] are recorded by mounting omnidirectional miniature microphones in earplugs of the mannequin's ears (Head and Torso Simulator (HATS) and Knowles Electronic Manikin for Acoustic Research (KEMAR) respectively). The same equipment and speech corpus (TIMIT) was used for the training of SONET, but it was done in anechoic conditions.

### 4.2. Dataset

SONET [18] was trained on the TIMIT speech corpus [20]. However, its training was done in the anechoic conditions (i.e. $RT_{60}$ of 0ms of [4]). When it is recycled for the dereverberation task, the test dataset comprises 5 speech samples of 5 different speakers (3 female and 2 male speakers). Although, these 5 speech samples were also taken from the TIMIT corpus, they were not used for the training of SONET. The duration of each clean speech is 1.05 seconds.

A reverberant speech is generated from the clean speech sample by first normalizing and then convolving it with the binaural room impulse responses (BRIRs), according to its position inside the room. After convolution with the reverberant room's RIR, the duration of speech signal exceeds beyond 1.05 seconds, but the signal is not clipped back to 1.05 seconds to keep the reverberations intact. The signal at both the microphones of binaural setup is then converted to TF domain by eq. (2), and its ILD and IPD spectrograms, calculated by equations (3) and (4) are given to the pre-trained ILD and IPD SONETs, and their outputs are combined by eq. (5), to extract the TF points of the direct path signal. The extracted direct path signal in TF domain is then converted to time domain by inverse STFT (ISTFT) and evaluated against the clean signal.

This process is then repeated for each of the five speech sources at each of the six position of figure 3 and the results of all positions are averaged within the room and then over all the five rooms. So, the results shown in Figure 4 are obtained by averaging the results of 150 speech samples (5 (speech signals) × 6 (positions) × 5 rooms).

### 4.3. Binaural room impulse responses (BRIRs)

We have not recorded the BRIRs ourselves, but have taken them from [4] and [5]. These are the real room impulse responses. These impulse responses are selected as they cover nearly all kinds of reverberant conditions that exist in most of the real-world situations. The acoustic properties of these rooms are given in table 1 below.

**Table 1:** Characteristics of different rooms used for experiments

| Room | A | B | C | D | S |
|---|---|---|---|---|---|
| $RT_{60}$ (ms) | 320 | 470 | 680 | 890 | 560 (Centre of room) |
| Room dimensions* | 6.6 × 5.7 × 2.3 | 4.6 × 4.6 × 2.6 | 18.8× 23.5 × 4.6 | 8.7 × 8.0 × 4.25 | 5 × 9 × 3.5 |

*Length × Width × Height (all in m)*

The source to microphone setup spacing is 1 m in room S, while it is 1.5 m in all other rooms. The details of each room can be found in [4] and [5].

The reverberation time is usually longer as the room volume increases [21]. According to this criteria, room C seems to be an odd one, as its volume is the largest of all rooms, and so, its $RT_{60}$ should be the largest of all. But the reason for not being so is because room C is actually a large cinema–style lecture theatre that seats 418 people. The abundance of soft seating and the low ceiling of the area around the lectern resulted in a relatively small $RT_{60}$ for the size of room C [4]. The $RT_{60}$ of room, in addition to its volume, depends on the absorption coefficient of walls and the objects inside the room, shape of the room, audience and the type of seating [22].

### *4.4. Pre-Trained Network*

Although there are many SONET versions according to the azimuth separation between the two sources, mentioned in [18], the best dereverberation results were obtained with SONET ($90^0$) for all the positions shown in figure 3.

### *4.5. Evaluation metrics*

Five evaluation metrics are used for comparison of different algorithms. These are signal to distortion ratio (SDR) [23], short term objective intelligibility (STOI) [24], signal to reverberation energy ratio (SRMR) [25], perceptual evaluation of speech quality (PESQ) [26] and cepstral distance (CD) [27]. Except for CD, higher is better for all metrics. SRMR does not require any reference, while all other metrics use the direct path signal to the left microphone as a reference for the performance evaluation.

### *4.6. Comparison algorithms*

We have compared our proposed algorithm in five different rooms stated in table 1 with five algorithms. These are spectral subtraction (SS) [12], spectral magnitude inverse filtering (SMIF) [1], weighted prediction error (WPE) [13], the precedence effect (PE) [14], and the orthogonal matching pursuit with spectral subtraction (OSS) [15]. A brief overview of these algorithms is already given in section 1.

We have tested the first four algorithms under the acoustic conditions of table 1. But due to limited computational resources available to us, we have mentioned the results of the last algorithm ([15]) directly from its paper, along with the acoustic conditions under which the experiments were carried out. The model of [15] has also been trained on the TIMIT dataset.

Fair comparison of D-S with other deep learning algorithms (e.g. [2] or [16]) is not possible, even from their respective papers due to the differences in their acoustic conditions and datasets. Also, the SEGAN based dereverberation model of [16] has only reported its results in terms of CER for the ASR application, while our proposed algorithm D-S is just a speech enhancement model, which enhances speech in presence of reverberations. ASR is currently out of its scope. In future, retraining [2] and [16] under similar acoustic conditions and testing them with the same dataset, as we have used for our proposed model, may decide the better performer.

### 4.7. Short time Fourier transform (STFT) Parameters

The STFT parameters used for converting time domain signal to time frequency domain are summarized in table 2 below.

**Table 2:** STFT parameters

| Sampling frequency | 16 kHz |
|---|---|
| Window Shape | Hamming |
| STFT frame length | 1024 samples |
| Hop size | 256 samples |

## 5. Experimentation Results and Comparison

The average results of different algorithms in all the five rooms of table 1 are shown in figure 4 under the noiseless conditions.

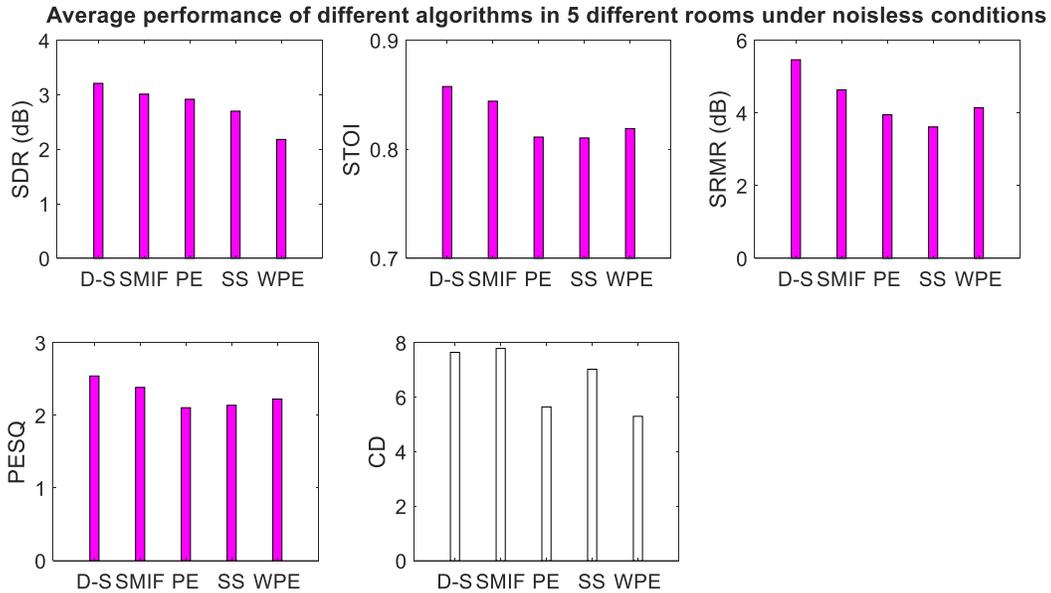

**Figure 4:** Performance of different algorithms in room A, B, C, D and S. The dark shaded bars show the metrics where our proposed algorithm (D-S) performs better than others, while the light shaded bars show the metrics for which our proposed model does not perform so well.

As clear from figure 4, our proposed dereverb algorithm performs better than the competing algorithms on most of the evaluation metrics. If we compare the average performance of other systems with our proposed algorithm, we can see that the deep learning of D-S has outperformed the classical signal processing models at every performance criteria except at CD. The SMIF dereverb model occupies the second position followed by other models. The improvement in SDR by D-S over SMIF is 0.2dB, in STOI 1.3%, in SRMR 0.83 dB, in PESQ by 0.16 points and in CD by 0.15 points. Although WPE and PE do not perform well at any of the metrics, yet they surpass other algorithms in case of CD, where the CD of PE is lower by 2 points and the CD of WPE is lower by 2.34 points than our proposed dereverb algorithm D-S.

Although the underlying cause of lower CD in figure 4 needs further investigation but it seems that this may be due to timbre distortion [23] caused by applying spectral masking. As the machine listening is based on template matching, so any possible degradation of even single frequency would affect the whole template [28], making our system unsuitable for ASR and ASV applications. However, the proposed model can be used for designing better hearing aids for hearing impaired persons and for assisted listening in reverberant and noisy conditions for non-native and normal hearing listeners.

The results of OSS as mentioned for simulated data in [15] are compared with D-S in table 3 below. The bold text shows the better of the two results in the comparison process.

**Table 3:** Performance comparison of OSS with D-S

| Method | $RT_{60}$ (s) | Data-set | PESQ | STOI | SRMR (dB) |
|---|---|---|---|---|---|
| OSS | 0.3 | TIMIT | 2.36 | 0.76 | **7.65** |
| D-S | 0.32 | TIMIT | **2.7** | **0.9** | 6.6 |

The results of OSS were evaluated at $RT_{60}$ of 0.3s which is lower than the $RT_{60}$ of our room A ($RT_{60}$ = 0.32s), yet our proposed model outperforms OSS at PESQ and STOI. However, the SRMR of D-S is 1.05 dB lower than OSS.

## 6. Conclusion

Our proposed dereverberation algorithm is the recycling of the existing neural network based speech separation algorithm SONET [18] for dereverberation of a single speech source. SONET has proved itself as a 'two in one' model. It can be used both for source separation and the dereverberation tasks. Although SONET was only trained in anechoic conditions ($RT_{60}$ = 0ms) with fixed target-microphone spacing of 1.5 m, yet it has performed well, the dereverberation task, for the unseen target-microphone spacing (1m) and for the unseen acoustic conditions ($RT_{60}$ > 0ms). However, one thing that must be kept is mind is that the SONET source separation model was trained on the interaural cues of two directional sources, whereas the reverberations (especially the late reflections) have a diffuse nature [19], so the chances were high, that it may not be able to recognize accurately the interaural cues of reflections coming from the directions on which SONET was not trained. But, due to the high learning capability of neural network, which has learnt the direct path interaural cues of target during the training phase of SONET, the pre-trained SONET source separation model has been able to satisfactorily recognize the direct path target cues and discriminate them from the unwanted reverberant cues.

Our proposed algorithm is good only for speech enhancement applications for human listening in reverberant conditions e.g. in designing better hearing aids or improving mobile speech quality when the talker is inside an enclosure. However, the proposed system is not recommended for forensic speech applications e.g. ASR or ASV due to its large CD as compared to other methods. This may be due to using spectral masking, which is found to be harmful for ASR applications [29]. In future, the use transfer learning on SONET may further improve the dereverberation results.

**Funding**

This project is funded by Higher Education Commission (HEC), Pakistan, under project no. 6330/KPK/NRPU/R&D/HEC/2016 and Higher Education Commission (HEC), Pakistan, under project no: 9827 with Ref no: KICS/UET/HRD/2019/36.